\documentclass[12pt,twoside]{article}

\usepackage{amsmath}
\input BoxedEPS
\SetRokickiEPSFSpecial  
\HideDisplacementBoxes

\makeatletter
\long\def\@makecaption#1#2{%
  \vskip\abovecaptionskip
  \sbox\@tempboxa{\small #1: #2}%
  \ifdim \wd\@tempboxa >\hsize
   \small #1: #2\par
  \else
    \global \@minipagefalse
    \hb@xt@\hsize{\hfil\box\@tempboxa\hfil}%
  \fi
  \vskip\belowcaptionskip}
\makeatother

\newcommand{\ket}[1]{\bigl|#1\bigr\rangle}
\newcommand{\zn}{z_1,z_2,\ldots,z_n}
\newcommand{\wn}{w_1,w_2,\ldots,w_n}
\newcommand{\npc}{NP-complete}
\newcommand{\Ht}{H(t)}
\newcommand{\Hn}[1]{H(#1)}
\newcommand{\pg}[1]{\ket{\psi_g(#1)}}
\newcommand{\ps}[1]{\ket{\psi(#1)}}
\newcommand{\rd}[1]{\mathop{\mathrm{d}#1}}

\newcommand{\sa}{satis\-fying assign\-ment}
\newcommand{\qa}{quantum algorithm}
\newcommand{\qaa}{quantum adiabatic algorithm}
\newcommand{\usa}{{\sc usa}}
\newcommand{\usi}{{\sc usa} instances}
\newcommand{\gusa}{{\sc gusa}}
\newcommand{\gusi}{{\sc gusa} instances}

\newcommand{\numeq}[2]{\begin{equation}
#2
\label{#1}
\end{equation}}
\newcommand{\refeq}[1]{(\ref{#1})}

\let\epsilon\varepsilon
\let\phi\varphi

\begin{document}
\title{A 
Quantum Adiabatic Evolution Algorithm\\ 
 Applied to Random Instances of\\
an NP-Complete Problem}
\author{Edward Farhi, Jeffrey
Goldstone\footnote{\tt farhi@mit.edu,
goldston@mit.edu}\\[-.75ex]
\footnotesize\itshape Center for Theoretical Physics, Massachusetts
Institute of Technology, Cambridge, MA 02139
\\ 
Sam Gutmann\footnote{\tt sgutm@neu.edu}\\[-.75ex]
\footnotesize\itshape  Department of Mathematics, Northeastern University,    
 Boston, MA 02115\\
Joshua Lapan, Andrew Lundgren,  Daniel Preda\footnote{\tt jlapan@mit.edu,
lundgren@mit.edu, pylot@mit.edu}\\[-.75ex]
\footnotesize\itshape  Massachusetts
Institute of Technology, Cambridge, MA 02139\\[-1.5ex]}
\date{\footnotesize\sf MIT-CTP~\#3035\\
A shorter version of this article appeared in the April 20, 2001 issue of
\emph{Science}.} 
\maketitle
\pagestyle{myheadings}
\markboth{\enspace E. Farhi, J. 
Goldstone,   S. Gutmann, J. Lapan, A. Lundgren,  D.
Preda}{A Quantum Adiabatic Evolution Algorithm Applied to 
Random Instances \ldots}
\vspace*{-2.25pc}

\begin {abstract}\noindent
A quantum system will stay near its instantaneous ground state if the Hamiltonian
that governs  its evolution varies slowly enough. This quantum adiabatic behavior is
the basis of a new class of algorithms  for quantum computing. We test one such
algorithm by applying it to randomly generated, hard, instances of an
\npc\ problem.
For the small examples that we can simulate, the \qaa\ works well,
and provides evidence that quantum computers (if large ones can be built) may be
able to outperform ordinary computers on hard sets of instances of \npc\ problems.
\end{abstract}

\vspace*{-1pc}

\setcounter{equation}{0}
\subsection{Introduction}
\label{sec1}

A large quantum computer has yet to be built, but the rules for programming such a
device, which are derived from the laws of quantum mechanics, are well established. 
It is already known that quantum computers could solve problems believed to be
intractable on classical computers.  (Throughout, ``classical" means non-quantum.)
An  intractable problem is one that necessarily takes too long to solve when the
input gets too big.   More precisely, a classically intractable problem is one that cannot
be solved using any classical algorithm whose running time grows only polynomially
as a function of the length of the input.  For example, all known classical factoring
algorithms require more than polynomial time as a function of the number of digits
in the integer  to be factored.  Shor's quantum algorithm for the factoring
problem~\cite{ref1} can factor an integer in a time that grows (roughly) as the square
of the number of digits.  This raises the question of whether quantum computers
could solve other classically difficult problems faster than classical computers.  

Beyond factoring there is a famous class of problems called NP-complete; see, for
example,~\cite{ref2new}.  Hundreds of problems are known to be in this class, ranging
from the practical (variants of the Traveling Salesman problem) to the whimsical (a
problem derived from the game of Go).  The original NP-complete problem,
Satisfiability, involves answering whether or not a Boolean formula (made up of
variables connected by the operators {\sc or}, {\sc and}, and {\sc not}) is true for
some choice of truth values for each variable. All NP-complete problems are related
in the following sense: if someone finds a polynomial-time algorithm for one
NP-complete problem, then this algorithm could be used as a subroutine in programs
that would then solve all other  NP-complete problems in polynomial time.  That no
one has succeeded in finding a classical polynomial-time algorithm for any of these
problems is strong evidence for the intractability of the whole class.  On the other
hand, no one has been able to prove that a polynomial-time algorithm cannot be
constructed for any NP-complete problem. 

It is  also an open question whether an
\npc\ problem could be solved in polynomial time on a quantum computer.  Here we
describe a recently developed approach to quantum computation based on quantum
adiabatic evolution~\cite{ref2}; for   related ideas, see~\cite{ref3new,ref4a}.  We
apply the \qaa\ to a specific NP-complete problem, Exact Cover (which is a restricted
form of Satisfiability).  A decisive mathematical analysis of this quantum  adiabatic
evolution algorithm has not been possible. Instead we resort to numerical
simulation of the running of the quantum algorithm~\cite{ref3}.  Each time we do the
simulation we use  as input a randomly generated instance of Exact Cover.  The
lengths of these inputs are necessarily small because simulating a quantum computer
on a classical computer requires memory that grows exponentially in the length of the
input.  On these small inputs our data looks promising. For our randomly generated
instances of Exact Cover we find that the quantum algorithm succeeds in a time that
grows only quadratically in the length of the input.

Saying that an algorithm solves a problem in polynomial time means that the
running time of the algorithm is bounded by a polynomial function of the length of
the input, and  also that the algorithm succeeds on \emph{every} possible
input.  On the other hand, an algorithm may succeed in polynomial time on a large set
of inputs but not on all.  (For many applications, this kind of almost-solution to the
problem may suffice.) This has led to efforts to identify sets of instances that are
hard for particular classical algorithms. In the last ten years, researchers working
on the \npc\ problem 3-SAT (another restriction  of Satisfiability) have identified a
set of instances that are hard for standard classical search algorithms~\cite{ref4}
(see also~\cite{ref7a} and references therein).  

While the quantum adiabatic
evolution algorithm could be applied to 3-SAT, we find it more convenient to study
Exact Cover.  Our instances of Exact Cover are generated from a set that we believe to
be classically intractable for large enough inputs.  (This set of Exact Cover instances
has similar properties to the set of 3-SAT instances that are identified as hard
in~\cite{ref4}.) Using a running time that grows only quadratically in the length of
the input, the quantum adiabatic algorithm solves the Exact Cover instances we
randomly generated.   Again, because of the space requirements inherent in
simulating a quantum computer, these instances are necessarily small.  Imagine that
this quadratic behavior actually persists for almost all arbitrarily large instances
generated from the same hard set.  This would not be enough to imply that the
quantum adiabatic algorithm solves an NP-complete problem because an algorithm
that solves a problem must work on \emph{all} instances.  However, if  classical
algorithms indeed require exponential time on this set and the quantum quadratic
behavior persists, then we would have found a new way in which quantum
computers could outperform classical computers.  

\subsection{Quantum adiabatic evolution as a computational tool}
\label{sec2}

All quantum systems evolve in time according to the Schr\"odinger equation 
\numeq{eq1}{
i \frac {\rd{}}{\rd t} \ps t
 = H(t) \ps t
}
where $\ps t$ is the time-dependent state vector and $\Ht$ is the
time-dependent Hamiltonian operator. (Here units have been chosen so that Planck's
constant~$\hbar$ equals~1.)
A quantum computer algorithm can be viewed as a specification of a Hamiltonian
$\Ht$ and an initial state $\ps 0$. These are chosen so that the state at time~$T$,
$\ps T$, encodes the answer to the problem at hand. 

In designing our quantum algorithm we take advantage of the quantum adiabatic
theorem, which we now explain.
 At time~$t$, the Hamiltonian $\Ht$ has an instantaneous
ground state $\pg t$, which is the eigenstate of $\Ht$ with the lowest energy.
Adiabatic evolution refers to the situation where $\Ht$ is slowly varying. Suppose
our quantum system starts at $t=0$ in the ground state of $\Hn0$, that is, $\pg0$.
The adiabatic theorem says that if $\Hn t$ varies slowly enough, then the evolving
state vector $\ps t$ will remain close to the instantaneous ground state $\pg
t$. (For a more precise discussion of the adiabatic theorem as it applies to our
situation, see~\cite{ref2}.)

To specify our algorithm we must give $\Ht$ for $0\le t\le T$, where~$T$ is the
running time of the algorithm.  We choose $\Ht$ so that the ground state of $\Hn 0$ is known in
advance and is easy to construct. 
For any instance of the problem under study, there is a Hamiltonian, $H_P$, whose
ground state encodes the solution. Although it is straightforward to construct $H_P$,
finding its ground state is computationally difficult. We take $\Hn T = H_P$, which
means that
$\pg T$ encodes the solution. For intermediate times,
$\Ht$  smoothly interpolates between
$\Hn0$ and $\Hn T = H_P$. We start with the quantum state in the known 
ground state of $\Hn0$. If the running time~$T$ is large enough, $\Hn t$ will indeed
be slowly varying, and by the adiabatic theorem  the final state reached, $\ps T$, will
be close to
$\pg T$, which encodes the solution to our problem.

\subsection{Exact Cover}
\label{sec3}

We focus on the \npc\ problem Exact Cover (see, for example, \cite{ref2new}).
Consider
$n$~bits
$z_1, z_2,
\ldots, z_n$ each of which can take the value~0 or~1. An $n$-bit instance of Exact
Cover is built up from clauses, each of which is a constraint imposed on the values of
three of the bits. If a given clause involves the three bits labeled $i$, $j$, and $k$,
then the  constraint is $z_i + z_j + z_k =1$,  which means that one of the three bits
must have the value~1 and the other two must have the value~0. Without the
constraint there are 8 possible assignments for the values of
$z_i$,
$z_j$, and $z_k$, but only 3 out of the 8 satisfy the constraint. 

An $n$-bit instance of Exact Cover is a list of triples $(i,j,k)$ indicating which groups
of three bits are involved in clauses. The problem is to determine whether or not
there is some assignment of the $n$~bit values that satisfies \emph{all} of the clauses.
Given an  assignment of values for $z_1, z_2, \ldots, z_n$ we can easily check
whether the assignment satisfies all of the clauses. But attempting to determine
whether or not at least one of the $2^n$ assignments  of  $z_1, z_2, \ldots,
z_n$ satisfies all the clauses is in fact an \npc\ problem. No known algorithm will
answer this question in general in a time that grows only as a polynomial in~$n$, the
number of bits. This dichotomy -- that it is easy to verify a solution if one is
presented but hard to determine if a solution exists -- typifies \npc\ problems. 

\subsection{Constructing the Hamiltonian}
\label{sec4}

To specify our quantum algorithm we must give the Hamiltonian $\Ht$ that
governs the evolution of the quantum system via \refeq{eq1}. The state vector
evolves in a Hilbert space of dimension~$2^n$. We take as a basis the $2^n$ vectors 
\numeq{eq2}{
\ket{z_1} \ket{z_2}\cdots \ket{z_n}
}
where each $z_i=0$ or~1. This $n$-qubit Hilbert space can be realized as a system
of $n$ spin-1/2 particles where $\ket{z_i=0}$ corresponds to the $i^{\textrm{th}}$
spin being up in the $z$-direction and $\ket{z_i=1}$ corresponds to  
 spin down in the $z$-direction. 

For $\Hn 0$ we couple a magnetic field in the $x$-direction  
to each quantum spin. 
(Specifically, the strength of the field at each site is equal to the number of clauses
that contain the bit. Thus  $\Hn 0$ is instance dependent; see~\cite{ref2}.)
The  ground state of the $i^{\textrm{th}}$ qubit
corresponding to  spin aligned in the $x$-direction is 
\numeq{eq3}{
\frac1{\sqrt2} \Bigl(\ket{z_i=0} + \ket{z_i=1}\Bigr)\ .
}
The ground state of $\Hn 0$ for the $n$-qubit quantum system is therefore
\numeq{eq4}{
\pg0  = \frac1{2^{n/2}} \sum \ket{z_1} \ket{z_2}\cdots \ket{z_n}
}
where the sum  is over all $2^n$ basis vectors. This means that $\pg0$, which we
take to be the starting state of our algorithm, is a uniform superposition of states
corresponding to all possible assignments of bit values. 

Recall that $\Hn T$ is to be chosen so that its ground state, $\pg T$, encodes the
solution to the instance of Exact Cover under consideration. To accomplish this we
first define a classical energy function $h(z_1, z_2, \ldots, z_n)$ that is a  sum of
energy functions $h_C (z_i,z_j,z_k)$ where $i$, $j$, and $k$ are the labels of the bits
involved in clause~$C$. It costs energy to violate clause~$C$: 
\numeq{eq5}{
h_C (z_i,z_j,z_k) = \begin{cases}
0 &\text{clause $C$ satisfied;}\\
1 &\text{clause $C$ violated.}
\end{cases}
}
Now let
\numeq{eq6}{
 h = \sum_C h_C
}
which means that the energy cost of any bit assignment $z_1, z_2, \ldots, z_n$ is
equal to the number of clauses that the assignment violates. We turn this classical
energy function into a quantum operator, diagonal in the $z$-basis:
\numeq{eq7}{
H_P \ket{z_1} \ket{z_2}\cdots \ket{z_n} = 
h(z_1, z_2, \ldots, z_n) \ket{z_1} \ket{z_2}\cdots \ket{z_n} \ .
}
This means that the ground state of $H_P$ corresponds to the bit assignment that
violates the minimal number of clauses. (If more than one assignment minimizes the
number of violations then there will be more than one ground state of $H_P$.) Recall
that the problem of Exact Cover is to determine if a given instance (specified by a set
of clauses) has a satisfying assignment. If we could use our quantum computer to
produce the ground state of $H_P$ we would then be able to determine if there is a
satisfying assignment. 

We define the time-dependent Hamiltonian by the linear interpolation
\numeq{eq8}{
\Ht = \Bigl(1-\frac tT\Bigr) \Hn0 + \frac tT H_P
}
so that the final Hamiltonian $\Hn T$ is $H_P$. Note that as~$T$ gets bigger $\Ht$
becomes more slowly varying and by the adiabatic theorem $\ps t$, which obeys
\refeq{eq1} and begins in $\pg0$, will stay close to $\pg t$. Therefore $\ps T$ will be
close to $\pg T$, the ground state of~$H_P$. Since the ground state of $H_P$
encodes the desired solution, a measurement (as described next) of the quantum state
at time~$T$ will with high probability determine if the instance has a satisfying
assignment. 

While \refeq{eq8} is the obvious path from $\Hn0$ to $H_P$, other paths are worth
exploring and may have algorithmic advantages. For example, any term of the form
$(\frac tT) (1-\frac tT) H_{\mathrm{extra}}$ can be added to $\Hn t$ and adiabatic
evolution will be maintained. We imagine $H_{\mathrm{extra}}$ as an
instance-independent Hamiltonian that is the sum of terms coupling only a few
qubits. In this paper, however, we only use the linear interpolation~\refeq{eq8}.

\subsection{Measurement}
\label{secmeasure}

At time T, we measure the $z$-component of each of the $n$ spins in the state
$\ps T$. This will produce a string $\zn$ of 0s and 1s that corresponds to one
of the $2^n$ bit assignments. This string can be quickly checked to see if it satisfies all
of the clauses.

For simplicity, suppose that the instance of Exact Cover has a satisfying assignment
and that it is unique, $\zn = \wn$. We can write
\numeq{eq9}{
\ps T = c_w \ket{w_1}\ket{w_2}\cdots\ket{w_n} +
\sum_z{}^{\raise1ex\hbox{$\scriptstyle\prime$}} c_z
\ket{z_1}\ket{z_2}\cdots\ket{z_n} }
where the sum $\sum'$ is over all $z=\zn$ other than $\wn$. The probability that the
measurement produces $\wn$ is $|c_w|^2$. 
Note that if we run the quantum algorithm again (with
the same instance, starting state $\ps 0$ and running time~$T$) we end up in
the same quantum state $\ps T$. However, the results of subsequent
measurements are all independent. The expected number of repetitions needed to
produce $\wn$ is $1/|c_w|^2$. 

\subsection{The running time \emph{T}}
\label{sec5}

The adiabatic theorem ensures that the quantum adiabatic evolution algorithm will
produce the desired state that encodes the solution to the instance of Exact Cover if
the running time~$T$ is long enough. 
This means that the success probability $|c_w|^2$, with $c_w$ given in \refeq{eq9},
goes to~1 as
$T$ goes to infinity.
 Determining how long is long enough to produce a reasonably large success
probability is the key to determining the potential usefulness of the algorithm. For
certain specialized examples we know that the required running time grows only as a
polynomial in the number of bits~\cite{ref2}. But addressing the general case of all
instances of Exact Cover is beyond our analytical abilities. In this paper we study
numerically the running time needed to solve a randomly generated set of Exact
Cover instances.

\subsection{Simulating a quantum computer}
\label{sec6}

The ingredients of the quantum adiabatic evolution algorithm are
\begin{enumerate}
\item[(i)] an instance-dependent, time-dependent Hamiltonian $\Ht$ given by
\refeq{eq8} 
\item[(ii)]  an initial state $\ps0= \pg 0$ given by \refeq{eq4}
\item[(iii)]  evolution from $0$ to $T$ according to 
the Schr\"odinger equation~\eqref{eq1}
\item[(iv)]  a measurement of $\ps T$. 
\end{enumerate}
In order to simulate this quantum computer we numerically integrate the
Schr\"odinger equation in (iii). For an $n$-qubit quantum system the state vector
$\ps t$ has $2^n$ complex components. For $n=20$  this means numerically
integrating a differential equation with 2,097,152 real variables. This is as large a
system as we could handle with our computer resources in a few months of running. 

Because the number of bits in our instances of Exact Cover is never more than 20, we
can always determine if the instance has satisfying assignments and what they are.
We do this by exhaustively checking, at 20 bits, all $2^{20}$ bit assignments, which
takes virtually no time on a classical computer. Admittedly we are only able to study
the performance of the quantum computer on instances that are easy for a classical
computer.  The classical simulation of a quantum computer is itself a classical
algorithm but not a good one. 

Ingredient~(iv) of the quantum algorithm is the measurement. However, in our
numerical simulation we do not simulate the measurement. Rather, since we have
already computed the $2^n$ coefficients of $\ps T$, we simply read off the
coefficients corresponding to \sa s, if there are any. The sum of the
squared magnitudes of these coefficients gives the probability that the actual
measurement would produce a satisfying assignment. 

\subsection{Generating random instances}
\label{secgenerating}

We begin by studying instances of Exact Cover with a unique satisfying assignment
(\usa). We believe that instances with only one satisfying assignment are the most  
difficult for the quantum algorithm. In fact, as we will see later, the
quantum algorithm runs faster on instances with more than one satisfying
assignment. We will also see that the \qa\  works well on  instances
with no \sa\ by producing an assignment that violates the minimal number of clauses.
Thus the restriction to a  \usa\ appears to restrict us to the most difficult cases
for the \qa. 

With the number of bits fixed to be $n$, we
generate \usa\ instances of Exact Cover as follows.  We pick
three distinct bits at random, uniformly over the integers
from $1$ to $n$.  
We then have a formula with one Exact Cover
clause.  We  add  clauses one at a time by picking  new sets of
three bits.  After each clause is added  we calculate the number of
satisfying assignments, which always decreases (or stays the same). If the number of
\sa s is reduced to just one, we stop and accept the instance.  If the number of \sa s
drops from more than one to zero without hitting one, we reject the instance and start
again.  Using this procedure, the number of clauses is not a fixed function of the
number of bits but rather varies from instance to instance.  

We have just described the procedure for randomly generating \usi\ on which we
test the \qa. Technically, we have thereby defined a probability distribution on the
set of \usi.  We refer to instances generated by this procedure as \gusi. The procedure
that generates \gusi\ does not produce each \usa\ instance with equal probability.
\gusi, on the average, have about as many clauses as bits, which is fewer clauses than
typical
\usi.

\subsection{The median time to achieve probability 1/8}
\label{secmedtime}

By the adiabatic theorem the probability of finding the \sa\ goes to~1 as the running
time goes to infinity. We are of course forced to settle for a finite running time and a
probability less than~1. We have (somewhat arbitrarily) picked a success probability
of~1/8 which, for $n\ge10$,  is much larger than $1/2^n$, the probability that a
random bit assignment is the \sa. For each number of bits~$n$ between 10 and 20 we
generate 75 \gusi\ of Exact Cover. For
each~$n$ we determine the median time required to achieve success probability
of~1/8. (Since this is a numerical study we actually hunt for a time that gives a
probability between 0.12 and 0.13.) 

\begin{figure}[hbt]
\centerline{\BoxedEPSF{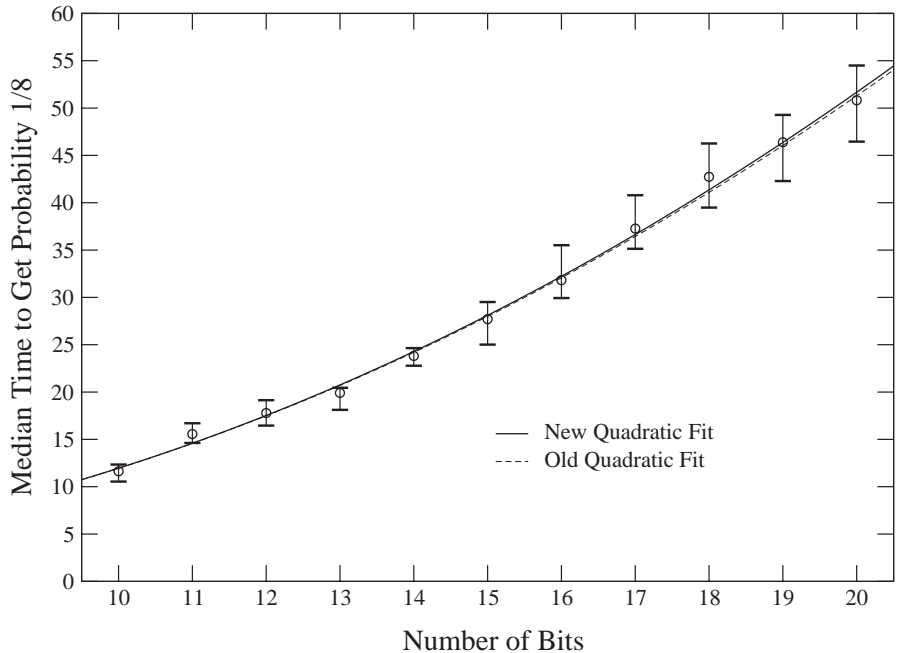}}
\caption{Each circle is the median time to achieve a success probability of
1/8 for 75 \gusi. The error bars give
95\% confidence limits for each median. The solid line is a quadratic fit to
the data. The broken line, which lies just below the solid line, is the quadratic fit
obtained in~\cite{ref3} for an independent data set up to 15 bits.}\label{fig1}
\end{figure}

In Figure~\ref{fig1} the circles represent the median time to achieve probability 1/8
for $10\le n \le 20$. The error bars give 95\% confidence limits on the medians. The
solid line is a quadratic fit to the data. In~\cite{ref3} corresponding data
was obtained for $7\le n\le15$. The dashed line in Figure~\ref{fig1} is the quadratic
fit to the data in~\cite{ref3}. 
The limited power of classical computers makes it impractical to go even a few
bits beyond 20, so further numerical study will not decisively determine how the
median running time grows with the number of bits. However, it is possible that the
data up to~20 bits already reveals the asymptotic performance of the algorithm. 

\subsection{Probabilities of success at a proposed running time}
\label{secprobsuc}

\begin{figure}[hbt]
\centerline{\BoxedEPSF{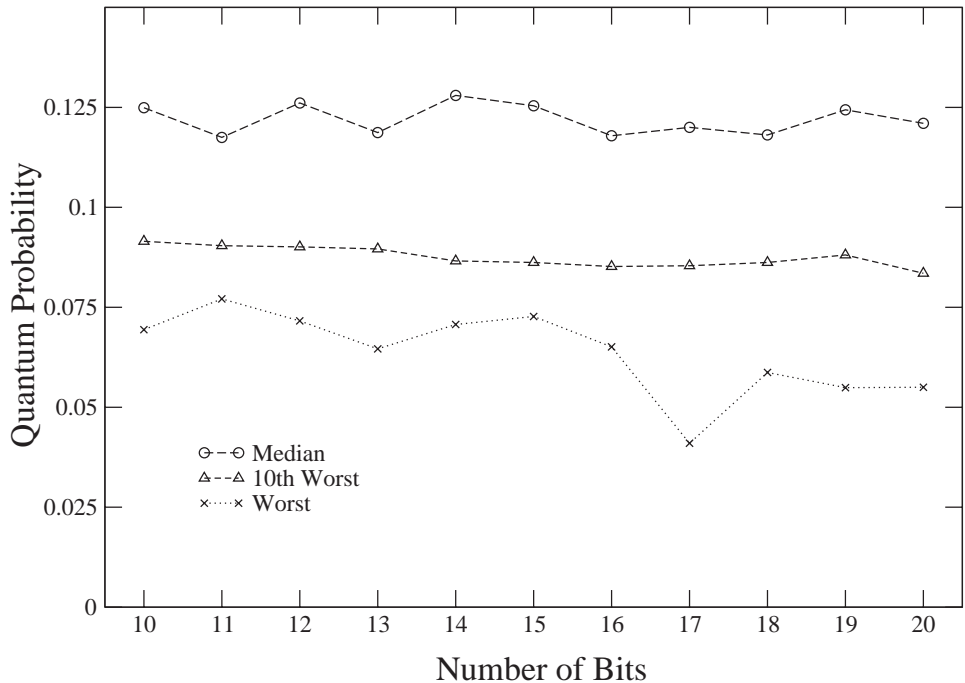 scaled 1600}}
\caption{Each circle is the median probability of success for 100 \gusi\
with the algorithm run at the solid line quadratic fit shown in
Figure~\ref{fig1}. Each triangle is the tenth-lowest probability and each
cross is the lowest probability. Note that the lowest probabilities do not
decrease much with~$n$.}\label{fig2}
\end{figure}

To implement the algorithm we want a running time that depends only on the
number of bits and not on the specific instance being considered.
 As a test we propose running
the quantum algorithm for a time
$T=T(n)$ that is equal to the quadratic fit to the median time required to achieve
probability 1/8, the solid curve shown in Figure~\ref{fig1}. For each $n$ between 10
and 20 we generate 100 new \gusi\ of Exact Cover and run the simulation on each
instance with $T=T(n)$. In Figure~\ref{fig2} the circles show the median probability of
success at each~$n$. 
Not surprisingly, these are all close to~1/8. We also show the tenth-worst and worst
probability for each~$n$. The good news for the \qa\ is that these do not appear to
decrease appreciably with~$n$. 

\begin{figure}[hbt]
\centerline{\BoxedEPSF{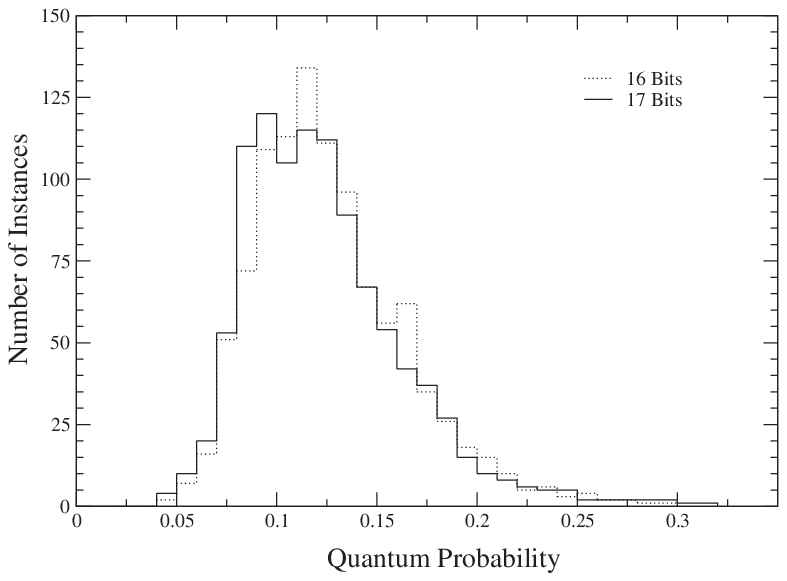 scaled 1600}}
\caption{Histograms of success probability for 1000 \gusi\ at 16 bits
and another 1000 at 17~bits. The running times are given by the solid
line quadratic fit in Figure~\ref{fig1} at $n=16$ and $n=17$.}\label{fig3}
\end{figure}

To further explore this we generate 1000 new \gusi\ of Exact Cover   
at both 16 and 17 bits. In Figure~\ref{fig3} we show the 
histograms of the success probability when the instances are run at $T(16)$ and
$T(17)$, respectively. The histograms indicate that a \gusa\ instance 
with success probability at or below~0.04 is very unlikely. 

If an algorithm (classical or quantum) succeeds with probability at least~$p$, then
running the algorithm $k$ times gives a success probability of at least $1-(1-p)^k$.
For example, if $p=0.04$, then 200 repetitions of the algorithm gives a success
probability of better than 0.9997. Suppose that as the number of bits increases it
remains true that almost all  \gusi\ have a success probability
of at least 0.04 at the quadratic run time $T(n)$. Then any $n$-independent desired
probability of success can be achieved with  a  fixed number of  repetitions.

\subsection{Instances with the number of clauses fixed in advance}
\label{secinstnum}

To study the performance of the \qa\ on instances that do not necessarily have a
\usa, we generate new instances now by fixing the number of clauses in
advance. Instances are generated with a fixed number of randomly chosen
clauses and then separated into two categories, those with at least one \sa\ and
those with none. Both for instances with and without \sa s we run the \qa\ at the
quadratic run time~$T(n)$.
\begin{figure}[hbt]
\centerline{\BoxedEPSF{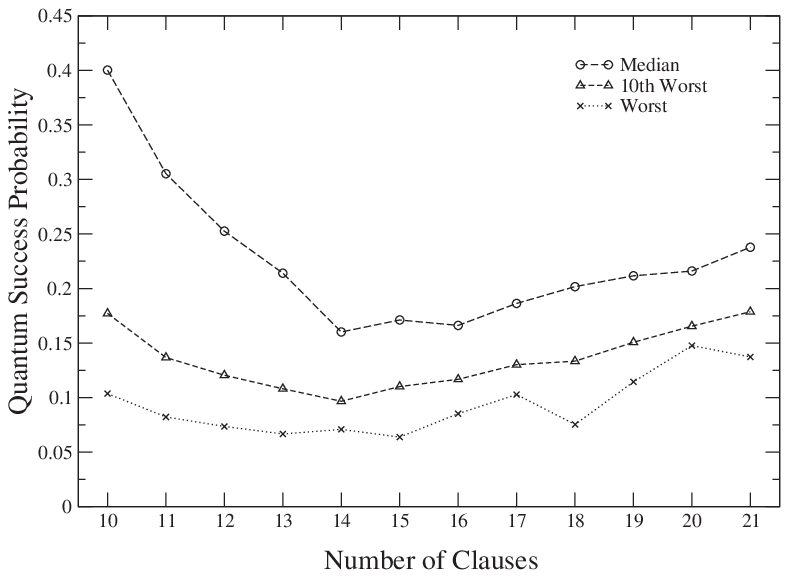 scaled 1600}}
\caption{Success probabilities for instances with one or more satisfying
assignments. Each circle is the median of 100 instances with the number of
clauses fixed. Each triangle is the tenth-worst probability and each cross
is the worst. All instances have 17 bits and the algorithm is run at the solid
line quadratic fit in Figure~\ref{fig1} at  $n=17$.}\label{fig4}
\end{figure}

In Figure~\ref{fig4} we show the success probability of the \qa\ at 17 bits for
instances with one or more \sa s. The number of clauses varies from 10 to 21 and at
each number of clauses there are 100 instances. The circles show the median
probability of success at each number of clauses. Note that these medians are well
above~1/8. Also note that the worst probabilities are never below~0.04. Very similar
results were obtained at 16 bits. Thus using the data from \gusi\  gives
a running time that is apparently adequate for all instances with \sa s. Instances
with a \usa\ appear to be  more difficult for our
algorithm than instances with multiple \sa s.

If an instance has no \sa\ then the minimum of the energy cost~\refeq{eq6} occurs at
those bit assignments that violate the fewest number of clauses. Correspondingly,
quantum adiabatic evolution leads to a quantum state that encodes these
assignments (in the limit that~$T$ goes to infinity). We say that the \qa\ succeeds if
after the quadratic run time $T(n)$, a measurement produces a bit assignment that
has the minimum number of violated clauses.

\begin{figure}[hbt]
\centerline{\BoxedEPSF{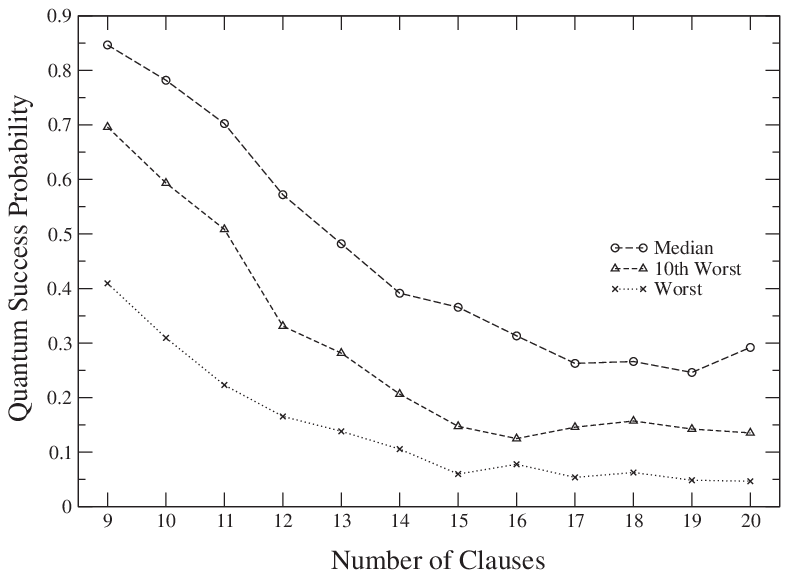 scaled 1600}}
\caption{Probability of obtaining an assignment violating the minimal
number of clauses for instances with no satisfying assignment. Each circle
is the median of 100 instances with the number of clauses fixed. Each
triangle is the tenth-worst probability and each cross is the worst. All
instances have 16 bits and the algorithm is run at the solid line quadratic
fit in Figure~\ref{fig1} at  $n=16$.}
\label{fig5}
\end{figure}

In Figure~\ref{fig5} we show the success probability of the \qa\ at 16~bits for
instances with no \sa. There are 100 instances at each number of clauses
from 9 to~20. The median probability of success is consistently above 1/8 and the
worst probability is never below 0.04. Similar results were established at 17~bits. 

The \qaa\ is designed to find the ground state of $H_P$, which corresponds to the
minimum of the classical energy cost~$h$, whether the instance has \sa s or not. The
data in Figures~\ref{fig4} and~\ref{fig5} show that the algorithm does at least as well
on the general problem of minimizing the number of unsatisfied clauses as it does on
finding a \usa.

\subsection{Classically difficult sets of instances}
\label{secclassdiff}

In this section we present evidence that \usi\  are closely related to sets
of Satisfiability instances that recent empirical work has shown to be hard for
classical algorithms.

Empirical studies of the performance of classical algorithms on  3-SAT have 
focused on finding the number of clauses, as a function of the number of bits, that
makes randomly generated instances hard to solve. (Again, solving an instance means
determining if it has at least one
\sa.) For 3-SAT there is  a ``phase transition''
(\cite{ref5}; also see~\cite{ref7a}). For a fixed number of bits, as the number of
clauses increases,  the probability that a randomly generated instance will have a \sa\
decreases abruptly.  If the ratio of the number of clauses to the number of bits is~$c$,
then there is a number
$\alpha\approx 4.25$ such that for
$c<\alpha$ almost all instances have \sa s, and for $c>\alpha$ almost none do. At
$c=\alpha$,  about half of the instances have \sa s, and this is where the hard
instances are believed to be found. Though these studies are necessarily carried out
using particular classical algorithms, it is believed that the easy/hard distinction is
independent of the algorithm used. 
\begin{figure}[hbt]
\centerline{\BoxedEPSF{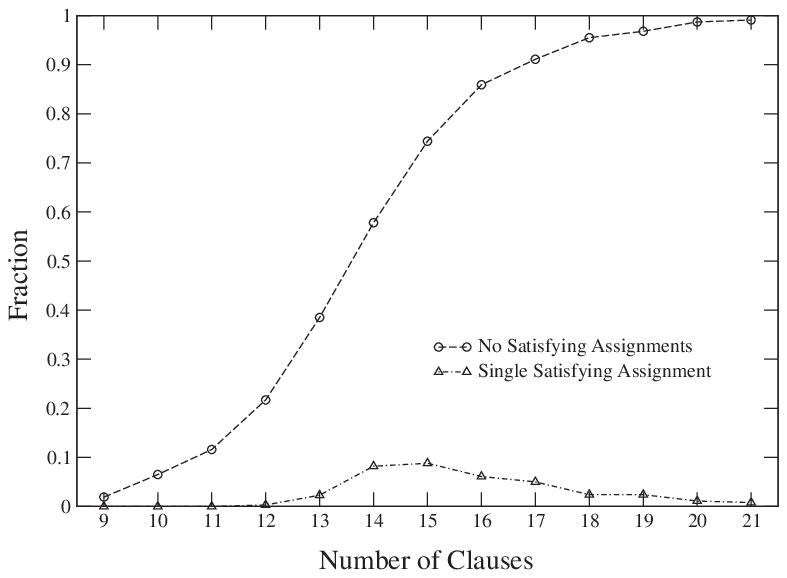 scaled 1677}}
\caption{The circles give the fraction of instances with no satisfying
assignment as a function of the number of clauses at 17~bits for Exact Cover. The
triangles give the fraction of instances at each number of clauses that have a unique
satisfying assignment.}\label{fig6}
\end{figure}

Even at our small number of bits, we begin to see phase transition behavior for
Exact Cover instances. The circles in Figure~\ref{fig6} are for 17 bits and show the
fraction of Exact Cover instances with \emph{no} \sa\ as a function of the number of
clauses. The steep rise is characteristic of the phase transition. The triangles show the
fraction of
all instances at a given number of clauses that are \usi. Importantly, this is peaked in
the hard region where the curve representing the fraction of unsatisfied instances is
rising from $\approx0$ to~$\approx1$. Thus we have some empirical support for the
intuitive notion that \usi\ will be hard for classical algorithms.  (\usi\ are not usually
used in numerical studies of classical algorithms because for higher numbers of bits
there is no simple way to generate instances known to be in the \usa\ class.)  


When  restricted to instances with no more than 20 bits, for which we are able to
simulate the \qa, classical search algorithms like GSAT~\cite{ref8} perform well on
\usi, or on instances with the number of clauses selected in the phase transition
region. 
In fact we ran GSAT on \gusi\ up to 20~bits and saw that the run time grows rather
slowly as the number of bits increases.
 It is believed that this behavior does not
persist, and that for larger~$n$ the time required by GSAT on the average to solve
instances randomly generated from  hard sets grows exponentially in the number of
bits~$n$. 

Since we cannot run quantum simulations much past $n=20$, we cannot tell whether
a similar fate befalls the \qa. However, it is possible that the asymptotic behavior of
the \qa\ is already visible in the performance up to $n=20$. It is conceivable that we
can see the large-$n$ behavior of the algorithm as soon as $2^n$ (rather than~$n$) is
large.

\subsection{Instances with structure}
\label{secinstruc}

In this article our focus is on randomly generated instances of Exact Cover.
In~\cite{ref2} quantum adiabatic evolution was studied analytically on certain
sequences of instances of Satisfiability where the clauses involve at most two bits.
Each of these sequences has enough structure to make it possible to determine the
required running time for any number of bits. For each case considered the \qaa\
succeeds in a time that grows only polynomially in the number of bits. Of course, the
structure of those instances makes it possible to determine the \sa\ by inspection, so
these instances are certainly easy for some classical algorithms. However, it is
encouraging that in the few cases where it is possible to analyze the \qa\
asymptotically it does perform well.

In~\cite{ref2} and~\cite{ref3} the \qaa\ was also applied to the problem of
unstructured search~\cite{ref9}. This problem can be cast as a restricted form of
Satisfiability where each instance has a single clause that involves all of the bits and
determines a unique
\sa. In this case the required running time \emph{must} grow exponentially in the
number of bits~\cite{ref10}. This exponential behavior is already clearly seen in the
data out to 14 bits in the numerical simulation of quantum adiabatic evolution
presented in~\cite{ref3}. 

We have also been looking for a structured sequence of instances of Exact Cover that
may be difficult for the \qaa. We now have a candidate sequence where the bits can
be viewed as arranged in a line with clauses connecting close-by bits. The success
probabilities drop sharply as a function of the number of bits when the algorithm is
run at the quadratic fit shown in Figure~\ref{fig1}. 
We have also experimented with \emph{ad hoc} modifications of the \qa\ that do
increase the success probability for this sequence.
These involve adding a term to $\Hn t$ as described at the end of Section~\ref{sec4}.
The $H_{\mathrm{extra}}$ used involves random couplings between the qubits.  In
any case,  sequences of structured instances have little bearing on the performance of
the
\qa\ on randomly generated sets, but are relevant to discussions of whether this
algorithm (or a modified version) could solve an \npc\ problem outright. Whether or
not a
\qaa\ could solve an \npc\ problem remains an open question.

\subsection{A novel approach to quantum computing}
\label{secnovappr}

The quantum adiabatic evolution algorithm operates in continuous time by evolving a
quantum state according to the Schr\"odinger equation~\refeq{eq1}. In the
conventional quantum computing paradigm an algorithm consists of a sequence of
discrete unitary transformations. Although the adiabatic time evolution can be well
approximated by a sequence of discrete unitary steps we see no advantage in this
reformulation. In fact, continuous time evolution may offer an alternative model for
the design of a quantum computer.

Quantum computation by adiabatic evolution works by keeping the quantum state
close to the instantaneous ground state of the Hamiltonian that governs the evolution.
This suggests that a device running the \qaa\ should be kept at a low temperature in
order to reduce unwanted transitions out of the ground state. 
Conventional quantum computing does not take place in the ground state and
decohering transitions caused by interactions with the environment are a major
impediment to current efforts to build a large-scale quantum computer. The
\qaa\ running on a cold device may be more fault tolerant than the implementations
of discrete step quantum computation usually envisioned.

Quantum computation by adiabatic evolution applied to a Satisfiability problem will
succeed if the running time is long enough. This follows from the quantum adiabatic
theorem. The \qaa\ is in a form that allows it to be applied to a wide variety of
combinatorial search problems; for example, see~\cite{ref11}. We have seen evidence
that for our randomly generated small instances of Exact Cover the required running
time grows slowly as a function of the number of bits. It is possible that the slow
growth we have already seen indicates the true asymptotic behavior of the algorithm
when applied to randomly generated hard instances of Exact Cover.
If this \qaa\ does actually outperform known classical algorithms, we would
have reason to eagerly await the construction of a quantum computer capable of
running it.

\subsection*{Acknowledgments}
This work was supported in
part by the Department of Energy under cooperative
agreement DE--FC02--94ER40818 and also by MIT's Undergraduate Research
Opportunities Program. We thank the MIT Laboratory for Nuclear Science's Computer
Facility for extensive use of the computer Abacus. We benefited greatly  from
conversations with Andrew Childs, Bart Selman, and Leslie Valiant. We also thank
Daniel Fisher, Bert Halperin, Mehran Kardar, and Patrick Lee for useful discussions
about the connection of our work to statistical mechanics systems.

\end{document}